\documentclass[aps,pre,twocolumn,showpacs,floatfix]{revtex4}
\usepackage{graphicx,bm,amssymb,amsmath,dcolumn,}
\usepackage{epstopdf}   
\usepackage{color}
\begin{document}
\newcommand{\be}{\begin{equation}}
\newcommand{\ee}{\end{equation}}
\newcommand{\half}{\frac{1}{2}}
\newcommand{\ith}{^{(i)}}
\newcommand{\im}{^{(i-1)}}
\newcommand{\gae}
{\,\hbox{\lower0.5ex\hbox{$\sim$}\llap{\raise0.5ex\hbox{$>$}}}\,}
\newcommand{\lae}
{\,\hbox{\lower0.5ex\hbox{$\sim$}\llap{\raise0.5ex\hbox{$<$}}}\,}


\definecolor{blue}{rgb}{0,0,1}
\definecolor{red}{rgb}{1,0,0}
\definecolor{green}{rgb}{0,1,0}
\newcommand{\blue}[1]{\textcolor{blue}{#1}}
\newcommand{\red}[1]{\textcolor{red}{#1}}
\newcommand{\green}[1]{\textcolor{green}{#1}}

\newcommand{\scrA}{{\mathcal A}} 
\newcommand{\scrN}{{\mathcal N}} 
\newcommand{\scrP}{{\mathcal P}} 
\newcommand{\scrQ}{{\mathcal Q}} 
\newcommand{\scrR}{{\mathcal R}}

\title{Crossover from isotropic to directed percolation}
\author{Zongzheng Zhou$^{1}$, Ji Yang$^{1}$, Robert M. Ziff$^2$ 
 $\footnote{Email: rziff@umich.edu}$, 
 Youjin Deng$^1$ 
 $\footnote{Email: yjdeng@ustc.edu.cn}$}
\affiliation{$^{1}$Hefei National Laboratory for Physical Sciences
at Microscale and Department of Modern Physics,
University of Science and Technology of China, Hefei, Anhui 230027, PR China }
\affiliation{{$^2$}Michigan Center for Theoretical Physics and Department of Chemical
Engineering, University of Michigan, Ann Arbor, Michigan 48109-2136, USA}

\date{\today} 
\begin{abstract}
  We generalize the directed percolation (DP) model by relaxing the strict directionality of DP
  such that propagation can occur in either direction but with anisotropic probabilities. We denote
  the probabilities as $p_{\downarrow}= p \cdot p_d$ and $p_{\uparrow}=p \cdot (1-p_d)$, 
  with $p $ representing the average occupation probability and $p_d$ controlling the anisotropy.
  The Leath-Alexandrowicz method is used to grow a cluster from an active seed site. 
  We call this model with two main growth directions {\em biased directed percolation} (BDP).
  Standard isotropic percolation (IP) and DP are the two limiting cases of the BDP model, 
  corresponding to $p_d=1/2$ and $p_d=0,1$ respectively.
  In this work, besides IP and DP, we also consider the $1/2<p_d<1$ region.
  Extensive Monte Carlo simulations are carried out on the square 
  and the simple-cubic lattices, and the numerical data are analyzed by
  finite-size scaling. We locate the percolation thresholds of the BDP model for $p_d=0.6$ and $0.8$, 
  and determine various critical exponents. 
  These exponents are found to be consistent with those for standard DP. 
  We also determine the renormalization exponent associated with the asymmetric 
  perturbation due to $p_d -1/2 \neq 0$ near IP, and confirm that such an asymmetric scaling field is relevant at IP.
\end{abstract}
\maketitle 

\section{Introduction}
  Directed percolation (DP), introduced in 1957 by Broadbent and Hammersley \cite{DP}, 
  is a fundamental model in non-equilibrium statistical mechanics and 
  represents the most common dynamic universality class \cite{Marro-Dickman}. 
  DP has a very wide application, including flow in a porous rock in a gravitational field, 
  forest fires, epidemic spreading, and surface chemical reactions~\cite{Grassberger-JSP-79-13}. 
  The DP process can be illustrated in the simple example of bond DP on the square lattice. 
  Along the horizontal (vertical) edges of the lattice, the propagation 
  occurs in a particular direction only, e.g., toward the right (the up).
  Frequently, the preferred spreading direction is termed ``temporal,'' 
  and the perpendicular one is called ``spatial;''
  the two-dimensional DP is thus often called ``(1+1)-dimensional DP.''
  The DP process has two distinct phases: the inactive phase for small occupation probability $p$ 
  where the propagation quickly dies out, and the active phase for large $p<1$.  
  Between these two phases, a transition occurs at $p_c$.
  As the threshold $p_c$ is approached, the temporal ($\parallel$) and the spatial ($\perp$) correlation lengths 
  diverge but with distinct critical exponents: $\xi_{\parallel}\sim|p-p_c|^{-\nu_{\parallel}}$ and
  $\xi_{\perp}\sim|p-p_c|^{-\nu_{\perp}}$. 
  The anisotropy is characterized by the so-called dynamic exponent $z=\nu_{\parallel}/\nu_{\perp}$. 
  For $p>p_c$, the order parameter $\scrP_{\infty}$, defined as the probability that 
  a randomly selected site can generate an infinite cluster, becomes non-zero and its behavior can be described 
  as $\scrP_{\infty}\sim (p-p_c)^{\beta}$, with $\beta$ another critical exponent.
  Below the upper critical dimensionality $(d_c+1)$ with $d_c=4$, 
  the three independent critical exponents, $\nu_{\parallel}$,
  $\beta$, and $z$, are sufficient to describe the DP universality class.
  While analytical results are scarce for DP, even in $(1+1)$ dimensions, 
  approximation techniques like series expansion \cite{SE1,SE2,SE3,SE4,SE5}
  and Monte Carlo simulations \cite{Grassberger-JPA-1989,Grassberger-Zhang,Lubeck,Voigt-Ziff}
  have produced fruitful results. 
  Moreover, after a great deal of efforts, experimental realization of the DP process 
  has been achieved  \cite{Takeuchi,errotum-Takeuchi,extension-Takeuchi} in nematic liquid crystals, 
  where the DP transition occurs between two turbulent states.

  Analogously, standard isotropic percolation (IP) \cite{IP} is a fundamental model in equilibrium statistical mechanics. 
  IP has attracted extensive research attention both in the physical and the mathematical communities, 
  and the critical behavior is now well understood. 
  Due to the isotropy, there exists only one spatial correlation length, 
  which scales as $\xi\sim|p-p_c|^{-\nu}$ near $p_c$. Numerous exact results are now 
  available in two dimensions (2D). For bond IP on the square lattice,
  the self-duality yields the threshold $p_c=1/2$~\cite{Kesten}; 
  the values of $p_c$ are also exactly known for bond and site percolation on several other 
  lattices~\cite{Ziff-06-0,Ziff-06-1,Ziff-10-0}, 
  or have been determined to a high precision~\cite{Feng08}.
  Thanks to conformal field theory and Coulomb gas theory~\cite{Cardy,Smirnov,Lawler,Kesten-1987}, the critical exponents 
  $\nu$ and $\beta$ are also exactly known as $\nu = 4/3$ and $\beta=5/36$.

  \begin{figure}
  \includegraphics[scale=0.5]{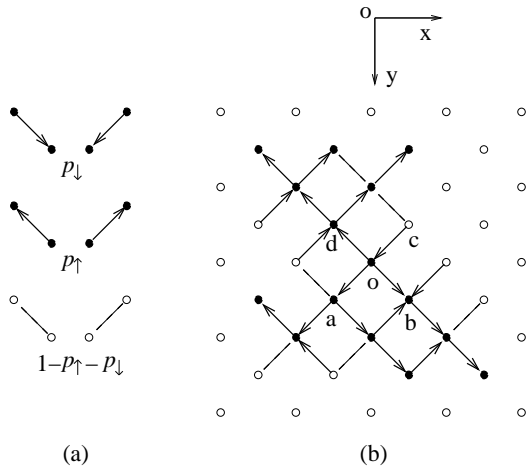}
  \centering
  \caption{ (a) State of an edge. (b) A typical cluster in the BDP process. 
  The seed is at site o, and the ``infected'' sites are denoted as solid dots.
  Dashed lines represent vacant bonds.}
  \label{conf}
  \end{figure}

  In this work, we introduce a generalized percolation propagation process that contains DP and 
  IP as two special cases.
  On a given lattice, each edge is assigned to one of the three possible states:
  occupied by a directed bond along a particular direction, 
  occupied by a directed bond against the particular direction, or unoccupied.
  This is illustrated in Fig.~\ref{conf} (a), and the associated probabilities are denoted as
  $p_{\downarrow}$, $p_{\uparrow}$, and $1-p_{\downarrow}-p_{\uparrow}$, respectively.
  As a result, the percolation process has two main growth directions. 
  For $p_{\downarrow}=p_{\uparrow}$, the symmetry between the two opposite directions 
  is restored, and the system reduces to standard bond IP. 
  In the limiting case $p_{\downarrow}=0$ or $1$, propagation 
  against or along the particular direction is forbidden, and 
  one has standard DP. We call this percolation model with two main growth
  directions {\em biased directed
  percolation} (BDP).
  We note that the BDP model is described by the field-theoretic equation in~\cite{Frey}.

  A natural question arises: in between standard DP and IP, what is the nature of 
  phase transition for BDP?
  For later convenience, we replace parameters  $p_{\downarrow}$ and $p_{\uparrow}$ by two new variables as
  \begin{equation}
  p_{\downarrow}=p \cdot p_d \; , \hspace{8mm}  p_{\uparrow}=p \cdot (1-p_d) \; .
  \label{eq_p0pd}
  \end{equation}
  The parameter $p$ is the average bond-occupation probability (irrespective of the bond direction),
  and $p_d$ accounts for the anisotropy.
  DP corresponds to $p_d=0$ or 1, while $p_d=1/2$ is for IP.

  In this work, extensive Monte Carlo simulations are carried out for BDP in two and three dimensions.
  A dimensionless ratio is defined to locate the percolation threshold.
  The data are analyzed by finite-size scaling, and the critical exponents are determined. 
  The numerical results suggest that the asymmetric perturbation due to $p_d -1/2 \neq 0$ 
  is relevant near IP, and thus that as long as $p_d \neq 1/2$, BDP is in the DP universality class. 
  These results further raise the following questions, remaining to be explored. 
  For IP, is the asymmetric renormalization exponent a ``new'' critical exponent or related in some way 
  to the known ones like $\nu$ and $\beta$? Particularly, can this ``new'' exponent be exactly obtained in two dimensions? 
  If so, what is the exact value?

  The remainder of this work is organized as follows. Section II introduces the BDP model, the sampled quantities,
  and the associated scaling behavior. Numerical results are presented in Secs. III and IV.
  A brief discussion is given in Sec. V.

\section{Model, sampled quantities, scaling behavior}

\subsection{Model}
  We shall describe in details the BDP model on the square lattice. 
  The generalization to higher dimensions is straightforward.

  As usual in the study of DP or IP, we view the BDP model as a stochastic growth process,
  and use the Leath-Alexandrowicz method~\cite{Leath,Alex} to grow the percolation cluster starting from a seed site. 
  Given the square lattice and the seed ``o'' in  Fig.~\ref{conf}(b), 
  for each of the neighboring edges of site o, a random number is drawn to determine the edge state. 
  If and only if the edge is occupied and the direction originates from the seed o, 
  the neighboring site is activated and belongs to the growing cluster. 
  For instance, in Fig.~\ref{conf}(b), the four neighboring edges of site o are all 
  occupied, but site c remains inactivated because of the ``wrong'' direction.
  After all the four neighboring edges have been visited, one continues 
  the growing procedure from the newly added sites. 
  In other words, one grows the percolation cluster shell by shell (the breadth-first scheme). 
  The growth of the cluster continues until the procedure dies out 
  or the maximum distance is reached, which is set at the beginning of the simulation. 

\subsection{Sampled quantities}

  In the cluster-growing process, the number of activated sites $N(s)$ is recorded 
  as a function of the shell number $s$. Let us count the shell of
  site ``o'' to be the first shell, the configuration in Fig.~\ref{conf}(b) 
  has $N=3, 5, 6$ for $s=2, 3, 4$, respectively. 
  Besides $N(s)$, one also records the Euclidean distance $r$ of each activated site 
  to the seed ``o'' for IP and to the $y$ axis for the anisotropic case.
  The reason for using different definitions of $r$ 
  is that, for the anisotropic case, 
  the average center of activated sites is expected to move linearly along the preferred direction,
  as $s$ increases. 
  Accordingly, we define a revised gyration radius $R(s)$ as 
  \begin{equation}
    \label{eq:gyration_radius}
    R(s) = \left\{ \begin{array} {ll}
                   0                              & \hspace{10mm} \mbox{if } N(s) = 0     \\
		   \sqrt{\sum_{i=1}^N r_i^2/N}    & \hspace{10mm} \mbox{if } N(s) \geq 1  \\
                   \end{array}
	    \right.
  \end{equation}
  The statistical averages $\scrN(s) \equiv \langle N(s) \rangle$ and  $\scrR(s) \equiv \langle R(s) \rangle$ 
  are then measured, as well as their statistical uncertainties.
  We also measure the survival probability $\scrP (s)$ that at least one site 
  remains activated at the $s$th shell and 
  the accumulated activated site number $\scrA (s) \equiv \langle\sum_{s'=1}^s N(s') \rangle $.

  In Monte Carlo study of critical phenomena and phase transitions, it is found that
  dimensionless ratios like the Binder cumulant are very useful in locating the critical point.
  Therefore, we also define a dimensionless ratio $ Q_{\scrN} (s) = \scrN(2s)/\scrN(s)$.

\subsection{Scaling behavior}

  Near the percolation threshold $p_c$, one expects the following scaling behavior
  \begin{eqnarray}
    \scrP(s,\epsilon) & \sim & s^{-Y_P} \mathbb{P} (\epsilon s^{Y_\epsilon}) \; ,  \nonumber \\
    \scrN(s,\epsilon) & \sim & s^{ Y_N} \mathbb{N} (\epsilon s^{Y_\epsilon}) \; ,  \nonumber \\
    \scrA(s,\epsilon) & \sim & s^{ Y_A} \mathbb{A} (\epsilon s^{Y_\epsilon}) \; ,  \nonumber \\
    \scrR(s,\epsilon) & \sim & s^{ Y_R} \mathbb{R} (\epsilon s^{Y_\epsilon}) \; ,  \nonumber \\
      Q_{\scrN}(s,\epsilon) & \sim & 2^{ Y_N} \mathbb{Q} (\epsilon s^{Y_\epsilon}) \; ,
   \label{eq:scaling}
  \end{eqnarray}
  where $\epsilon = p-p_c$ represents a small deviation from $p_c$.
  Symbols $Y_P, Y_N, Y_A, Y_R$, and $Y_\epsilon$ denote 
  the associated critical exponents, and $\mathbb{P}, \mathbb{N}, \mathbb{A}$,  $\mathbb{R}$, 
  and $\mathbb{Q}$ are universal functions.
  For simplicity, only one scaling field, 
  which accounts for the effect due to deviation from $p_c$, is explicitly included in Eq.~(\ref{eq:scaling}). 
  Right at $p_c$, as $s$ increases, the survival probability $\scrP(s)$ decays to zero
  while the other quantities diverge, except for the ratio $Q_\scrN$ which goes to a constant.
  A trivial relation is $Y_A = Y_N +1$.

  For standard DP ($p_d=0$ or 1), exponents $Y_P$ and $Y_N$ are normally denoted as $\delta$ and $\eta$, 
  respectively ($Y_N$ is also denoted as $\theta$ in \cite{Hinrichsen-2000}).
  It can be shown that exponent $Y_\epsilon$ is $Y_\epsilon = 1/\nu_\parallel$. Further, 
  exponent $Y_R$ relates to $\delta$ and the dynamic exponent $z$ as $Y_R = -\delta +1/z$,
  where $-\delta$ arises from the behavior $\scrP(s) \sim s^{-\delta}$.
  Below the upper critical dimensionality $(d_c+1)$ with $d_c=4$, there exist three independent exponents, 
  which can be chosen as $\nu_\parallel$, $\beta$, and $z$. The others can be obtained by the scaling
  relations \cite{Hinrichsen-2000}
  \begin{equation}
    \nu_\perp = \nu_\parallel/z \; , \hspace{5mm} \delta = \beta/\nu_\parallel \; , \hspace{5mm} 
    \eta = (d \nu_\perp - 2\beta)/\nu_\parallel \; ,
  \end{equation}
  where the last one involves the spatial dimensionality $d$ and is called the hyperscaling relation.
  In $(1+1)$ dimensions, these exponents have been determined to high precision: 
  $\nu_\parallel = 1.733\,847 (6)$, $\beta = 0.276\,486(8)$, and $z=1.580745(10)$ \cite{SE4}. 
  In $(2+1)$ dimensions, these exponents are $\nu_\parallel = 1.2890(7)$, $\beta = 0.581\,2(6)$,
  and $z=1.7665(4)$ \cite{Grassberger-Zhang,Voigt-Ziff,Perlsman,Junfeng}.

  For standard IP ($p_d=1/2$), the shell number $s$ is frequently called ``chemical distance'' \cite{Havlin},
  accounting for the minimum length among all the possible paths between the seed site and
  the activated sites on the $s$th shell. 
  At $p_c$, the length $s$ of the chemical path relates to the Euclidean distance $r$
  as $s \sim r^{d_{\rm min}}$ \cite{Grassberger-JPA-1992,Grassberger-JPA-1985}, with $d_{\rm min} \geq 1$ denoting the shortest-path exponent.
  In terms of the Euclidean distance $r$, it is known that the survival probability
  scales as $\scrP (r) \sim r^{-\beta/\nu} $, the accumulated site number $\scrA(r) \sim r^{\gamma/\nu}$,
  and the $p_c$-deviating scaling behavior $\epsilon r^{1/\nu}$.
  This immediately yields $Y_P = -\beta/(\nu d_{\rm min}), Y_A = \gamma/(\nu d_{\rm min}),
  Y_N = \gamma/(\nu d_{\rm min})-1, Y_R = (1-\beta/\nu)/d_{\rm min}$, and $Y_\epsilon = 1/(\nu d_{\rm min})$. 
  For IP, one has the scaling relation as
  \begin{equation}
  \gamma/\nu = d  -2\beta/\nu \; .
  \end{equation}
  In 2D, $\nu$ and $\beta$ are exactly known as $\nu = 4/3$ and $\beta = 5/36$, 
  which yield $\gamma/\nu = 43/24 \approx 1.79166\ldots$ 
  and $\beta/\nu = 5/48 \approx 0.104166\ldots$.
  The shortest-path exponent $d_{\rm min}$, together with the so-called backbone exponent, is 
    among the few critical exponents of which the exact values are not known for the 2D percolation 
    universality class.  It was conjectured to be $d_{\rm min} = 217/192 = 1.13020 \ldots$~\cite{Deng-PRE-81},
  and some recent estimates are 1.1306(3) \cite{Grassberger-JPA-32} and 1.13078(5) \cite{JY}.
  In three dimensions, no exact results are available, 
  and the numerical estimates are $d_{\rm min} =1.374(4)$~\cite{Grassberger-JPA-25}, $\beta/\nu=0.4774(1)$ and
  $\nu = 0.8734 (6)$~\cite{deng}, which yield $\beta = 0.4170 (4)$.

\section{Results}
  In this work, we consider the BDP model on the square lattice for 2D and the simple-cubic lattice for 3D.
  The simulation applies the aforementioned Leath-Alexandrowicz growth method. 
  The dimensionless ratio $Q_\scrN$ is used to locate the percolation threshold $p_c$. 
  According to Eq.~(\ref{eq:scaling}), ratio $Q_\scrN$
  is expected to have an approximate common intersection at $p_c$ for different shell number $s$.
  At the threshold $p_c$, as $s \rightarrow \infty$, the common intersection converges to a universal value $2^{Y_N}$
  and the slope of $Q_\scrN$ increases as $s^{Y_\epsilon}$.
 
\subsection{Standard IP}
\begin{figure}
  \includegraphics[scale=1.25]{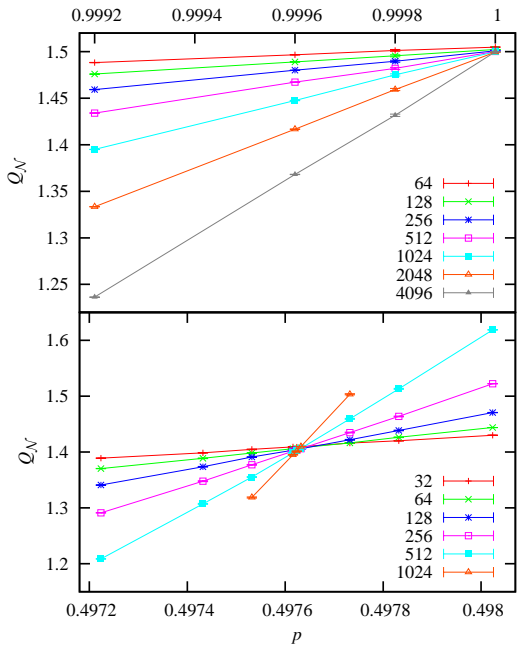}
  \centering
  \caption{Ratio $Q_\scrN$ for IP in 2D (top) and 3D (bottom).}  
  \label{fig:IP}
  \end{figure}

  Standard IP corresponds to $p_d=1/2$. Monte Carlo simulation 
  was carried out up to $s_{\rm max} = 8192$ for 2D and $2048$ for 3D. 
  About $10^8$ samples were taken for each data point on each lattice.
  The $Q_\scrN$ data are shown in Fig.~\ref{fig:IP}. Indeed, we find
  an approximate common intersection near $p=1$ and $0.4976$ for 2D
  and 3D, respectively. This agrees with the known threshold $p_c/2=1/2 \; (2D)$ 
  and $0.248\,812\,6 (5) \; (3D)$~\cite{Lorenz}. Note that, since the occupied bond
  can propagate the growth process only if it has the correct orientation, 
  there is a factor of 2 difference between the bond-occupation probability $p$ here and the $p$ of the equivalent bond percolation probability.

  To have a better estimate of $p_c$, according to a least-squared criterion, 
  the $Q_\scrN$ data are fitted by 
  \begin{eqnarray}
   && Q_\scrN(s,\epsilon)=Q_{\scrN,c} + \sum_{k=1}^4 q_k \epsilon^k s^{kY_\epsilon} +b_1 s^{y_1} \nonumber \\
   && +b_2 s^{-2} +c \epsilon s^{Y_\epsilon +y_1}+n \epsilon^2 s^{Y_\epsilon} + ...\; ,
  \label{eq:fitQ}
  \end{eqnarray}
  which is obtained by Taylor-expanding Eq.~(\ref{eq:scaling}) and taking into account 
  finite-size corrections due to the leading irrelevant scaling field and analytical background contribution.
  These are described by the two terms with amplitudes $b_1$ and $b_2$, of which
  the term with $n$ arises from the nonlinearity of the relevant scaling field in terms 
  of the deviation $\epsilon$, and the one with $c$ accounts for the combined effect of 
  the leading relevant and irrelevant scaling fields. In the fits, various formulas are tried, 
  which correspond to different combinations of those terms in Eq.~(\ref{eq:fitQ}). 
  For a given formula, the $Q_\scrN$ data for small $s <s_{\rm min}$ are gradually 
  excluded from the fits to see how the residual $\chi^2$ changes with respect to $s_{\rm min}$.
  The results from different formulas are compared with each other to estimate 
  the possible systematic errors. In two dimensions, we obtain $p_c = 1.000\, 000(4)$, 
  $Q_{\scrN,c} = 1.499\, 5(1)$, $Y_\epsilon = 0.664(3)$, and $y_1 = -0.96 (6)$. 
  Note that the leading irrelevant thermal scaling field is $ \omega = -2$ 
  for 2D percolation universality \cite{Ziff-PRE-2011}; apparently, the leading correction exponent $y_1 = -0.96$
  does not correspond to $\omega$. Instead, $y_1$ should be associated 
  with the chemical distance. From the relations $Q_{\scrN,c} = 2^{Y_N}$, $Y_N = \gamma/(\nu d_{\rm min}) -1$,
  and $Y_\epsilon = 1/ (\nu d_{\rm min})$, and the exact values $\gamma/\nu =43/24$ and $1/\nu = 3/4$,
  we determine $d_{\rm min} = 1.130\,76 (10)$ from $Q_{\scrN,c} = 1.499\, 5(1)$, 
  and $d_{\rm min} = 1.130 (6)$ from  $Y_\epsilon = 0.664(3)$.

  In three dimensions, our results are $p_c = 0.497\,624(1)$, $Q_{\scrN,c} = 1.400(1)$,
  $Y_\epsilon = 0.830(1)$, and $y_1 = -0.8(2)$. Our estimate of $p_c/2 = 0.248\,812\,0(5)$
  agrees with the existing one $0.248\,812\,6(5)$~\cite{Lorenz}, and has a
  comparable error margin. 

  \begin{table*}
  \begin{center}
  \begin{tabular}[t]{|ll|l|l|l|l|}
  \hline
         &          & $\beta$
                    & $\nu$
  	     	  & $d_{\rm min}$
	  	  & $p_c/2$ \\
  \hline
  $2D$ & (known)   & $5/36$~\cite{IP,Cardy,Smirnov,Lawler,Kesten-1987}               & $4/3$~\cite{IP,Cardy,Smirnov,Lawler,Kesten-1987}
                   & $1.130\,6 (3)$ ~\cite{Grassberger-JPA-32,Deng-PRE-81,JY}        & $1/2$~\cite{IP,Kesten}  \\
       & (present) & $0.138\,7(10)$        & $1.332(6)$   & $1.130\,76 (10)$    & $0.500\, 000(2)$ \\
  \hline
  $3D$ & (known)   & $0.4167(4)$          & $0.873\,4(6)$~\cite{deng}  
                                       & $1.374(4)$~\cite{Grassberger-JPA-25}  & $0.248\,812\,6(5)$~\cite{Lorenz} \\
        & (present) & $0.417(1)$       & $0.876(2)$      & $1.375(1)$          & $0.248\,812\,0(5)$ \\
  \hline
  \end{tabular}
  \caption{Percolation thresholds and critical exponents for IP.} 
  \label{tab:IP}
  \end{center}
  \end{table*}

  To estimate other critical exponents, we simulate right at
  the threshold $p/2 =1/2$ for 2D and $0.248\, 812 \, 0$ for 3D. The simulation was
  carried out for $s$ up to $s_{\rm max} =8192$ for 2D and 2048 for 3D.  Further, to eliminate one more unknown
  parameter in the fits, we measure the dimensionless ratios $Q_\scrP (s) = \scrP (2s)/\scrP(s)$ 
  and $Q_\scrR = \scrR (2s)/\scrR(s)$. These $Q$ data are fitted by
  \begin{equation}
    Q(s) = Q_c + b_1 s^{y_1} + b_2 s^{-2} \; .
  \label{eq:fitQc}
  \end{equation}
  In two dimensions, the results are $Q_{\scrP,c} = 0.9382(1)$ and $y_1 = -0.80(7)$ for $Q_\scrP$, and
  $Q_{\scrR,c} = 1.7318(2)$ and  $y_1 = -0.9(1)$ for $Q_\scrR$.
  For all these three ratios, the leading correction is described by an exponent $y_1 \approx -1$. 
  Taking into account the exact values $\beta/\nu = 5/48$, one has $d_{\rm min} = 1.132(2)$
  from $Q_{\scrP,c}$ and $d_{\rm min} = 1.130\, 7(3)$ from $Q_{\scrR,c}$.  

  In three dimensions, the results are $Q_{\scrP,c} = 0.7865(2)$ and $y_1 = -0.7(2)$ for $Q_\scrP$, and
  $Q_{\scrR,c} = 1.3020(3)$ and  $y_1 = -0.9(2)$ for $Q_\scrR$. 
  Combining the estimate $Q_{\scrP,c}$ and $Q_{\scrR,c}$ together, one has
  $\beta/\nu = 0.4765(8)$ and $d_{\rm min} = 1.375(1)$. Our result $d_{\rm min} = 1.375(1)$
  agrees well with the existing result $d_{\rm min} = 1.374(4)$~\cite{Grassberger-JPA-25}, 
  and significantly improves the error margin.

  For comparison, these results are summarized in Table~\ref{tab:IP}.

\subsection{Standard DP}
  We simulate standard DP by taking $p_d=1$. The simulation was carried out for $s$ up to $s_{\rm max} = 16384$ for 2D, 
  and 2048 for 3D. The number of samples for each data point is about $8 \times 10^8$ in 2D and $1.6 \times 10^8$ in 3D. 

\begin{figure}
  \includegraphics[scale=1.25]{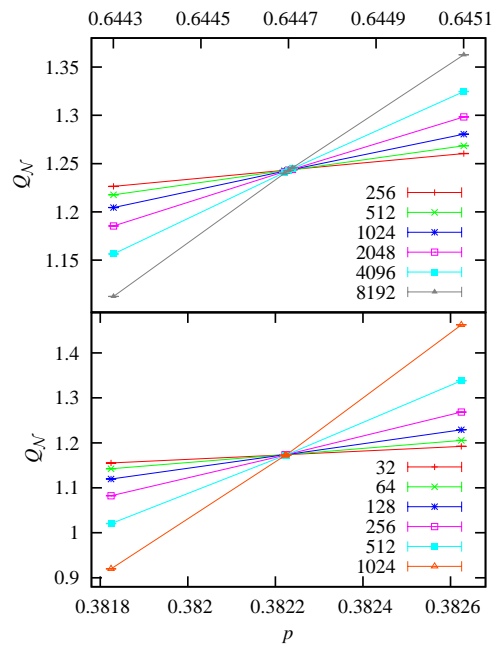}
  \centering
  \caption{Ratio $Q_\scrN$ for standard DP in 2D (top) and 3D (bottom).}
  \label{fig:DP}
  \end{figure}

  The $Q_\scrN$ data are shown in Fig.~\ref{fig:DP}. 
  A good intersection is observed for both 2D and 3D, which 
  yields $p_c = 0.64470 $ for 2D and $0.38222$ for 3D, from a rough visual fitting.
  We fit the $Q_\scrN$ data more precisely using Eq.~(\ref{eq:fitQ}). 
  On the square lattice, we obtain $p_c=0.644\,700\,5(8)$, 
  $Q_{\scrN,c}=1.242\, 9(2)$, $Y_\epsilon = 0.576(3)$, and $y_1=-0.9(1)$. 
  The estimate of the percolation threshold agrees well with the existing 
  more precise result $0.644\,700\,185 (5)$~\cite{SE4}.  
  From the relations $Q_{\scrN,c} = 2^{Y_N} = 2^\eta$ and $Y_\epsilon = 1/ \nu_\parallel$,
  we have $\eta = 0.313\,7 (2)$ and $\nu_\parallel = 1.736 (9)$.
  On the simple-cubic lattice, our results are $p_c=0.382\,225\,6 (5)$, $Y_\epsilon = 0.777(2)$,
  $Q_{\scrN,c}=1.1738(1)$, which yield $\eta = 0.2312(1)$ and $\nu_\parallel = 1.287(4)$.
  Here the $y_1$ is too small to estimate since the numerical data of $s\geq 24$ can be well
  described even though we do not include any corrections. The agreement of $p_c$ with the 
  existing estimate $p_c = 0.382\,224\,64 (4)$ \cite{Junfeng} is within two standard deviations. 

  Analogously, we simulate right at the percolation threshold $p_c = 0.644\,700\,185$ for 2D
  and $p_c = 0.382\,224\,64$ for 3D. The dimensionless ratios $Q_\scrP$
  and $Q_\scrR$ are measured, and the data are fitted by Eq.~(\ref{eq:fitQc}). 
  For 2D, the results are $Q_{\scrP,c} = 0.89537(5)$, $y_1 = -0.98(5)$ and $Q_{\scrR,c} = 1.3882(1)$,
  $y_1=-1.1(1)$, which yield $Y_P = \delta = 0.159\,44 (9)$ and
  $Y_{R} = (-\delta +1/z) = 0.47322 (10)$. Taking into account the estimates of
  $\nu_{\parallel}$ and $\delta$, one has $\nu_{\perp} = 1.098(6)$. For 3D, the results are
  $Q_{\scrP,c} = 0.7311(4)$ and $Q_{\scrR,c} = 1.0822(1)$, which yield that $\delta = 0.4519(8)$ and
  $\nu_{\perp} = 0.728(4)$.

  These results are listed in Table~\ref{tab:DP}.

\subsection{BDP}
  \begin{figure}
  \includegraphics[scale=1.25]{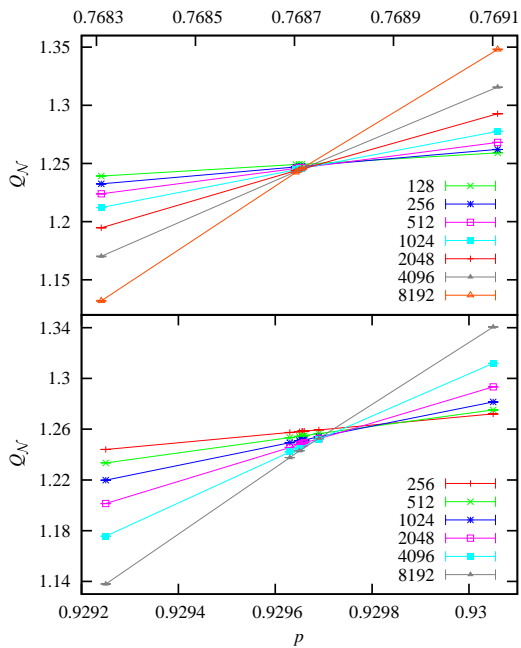}
  \centering
  \caption{Ratio $Q_\scrN$ for BDP in 2D. The top (bottom) panel corresponds to $p_d = 0.8$ ($0.6$) case.} 
  \label{fig:BDP2D}
  \end{figure}

  \begin{figure}
  \includegraphics[scale=1.25]{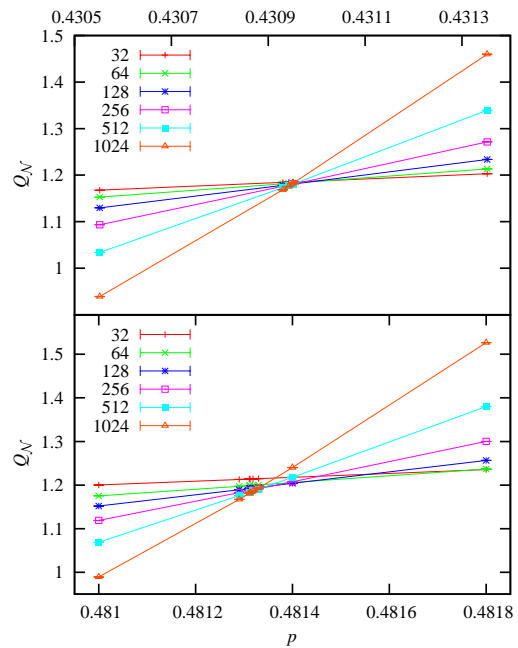} 
  \centering
  \caption{Ratio $Q_\scrN$ for BDP in 3D. The top (bottom) panel corresponds to $p_d = 0.8$ ($0.6$) case.} 
  \label{fig:BDP3D}
  \end{figure}

  For the purpose of studying BDP, we choose $p_d=0.6$ and $0.8$. 
  The simulation was carried out for $s$ up to $s_{\rm max} = 16384$ for 2D and 2048 for 3D. 
  About $2 \times 10^8 $ samples were taken for each data point in each case.

  The $Q_\scrN$ data are shown in Fig.~\ref{fig:BDP2D} for 2D and Fig.~\ref{fig:BDP3D} for 3D. 
  The transitions are also clearly observed, but the approximate common intersections
  are not as good as those for standard DP and IP. 
  This suggests the existence of additional finite-size corrections.

 The $Q_\scrN$ data are also fitted by Eq.~(\ref{eq:fitQ}) according to a least-squared criterion. 
 To account for the possible existence of additional corrections, we replace
 the terms in Eq.~(\ref{eq:fitQ}), with $b_1$, $b_2$, and $c$,
 by $b_is^{y_i}+b_1s^{y_1}+c \epsilon s^{y_i+y_\epsilon}$.
 The exponent $y_1$ is fixed at $-1$, in accordance 
 with our above estimate of $y_1$ for both standard IP and DP. 
 Indeed, the new source of finite-size correction can be identified in the fits, 
 which yield $y_i = -0.5 (2)$ both in 2D and 3D.
 The results for $p_c$, $\eta = \log_2 Q_{\scrN,c}$, and $\nu_{\parallel} = 1/Y_\epsilon$ 
 are summarized in Table~\ref{tab:DP}.

\begin{table*}
\begin{center}
{
\begin{tabular}[t]{|l|l|l|l|l|l|l|l|l|}
\hline
 D      & Ref.     & $p_d$   & $p_{c}$               & $\beta$            & $\nu_{\parallel}$ 
                       & $z$                   & $\eta$           & $\delta$          \\
\hline
2 &\cite{SE4}& 1       & $0.644\,700\,185(5)$  & $0.276\,486(8)$    & $1.733\,847(6)$ 
                       & $1.580\,745(10)$      & $0.313\,686(8)$    & $0.159\,464(6)$ \\
\hline
  &         & 1   & 0.644\,700\,5(8) & 0.277(2) & 1.736(9)&1.580\,6(3) &0.313\,7(2)  &0.159\,44(9) \\
  &         & 0.8 & 0.768\,708(1)    & 0.278(2) & 1.74(1) &1.577(5)    &0.314\,1(4)  &0.159\,5(1)   \\
  &         & 0.6 & 0.929\,668(3)    & 0.279(2) & 1.754(6)&1.578(5)    &0.316\,1(8)  &0.159(1)    \\
\hline
3 &\cite{Junfeng} & 1 & 0.382\,224\,64(4)   &0.581\,2(6)&1.289\,0(7) &1.766\,5(2)&0.230\,81(7) & 0.450\,9(2)\\
\hline
  &                & 1   & 0.382\,225\,6(5) &0.582(5)& 1.287(4) &1.767(3)&0.231\,2(1)&0.451\,9(8) \\
  &                  & 0.8 & 0.430\,941(2)  &0.577(5)& 1.289(5) & 1.77(1)&0.229(3) &0.448(2) \\
  &                  & 0.6 & 0.481\,310(2)  &0.583(8)& 1.292(5) & 1.76(2)&0.226(9) &0.452(4) \\
\hline
\end{tabular}
}
\caption{Percolation thresholds and critical exponents for standard DP ($p_d = 1$)  and BDP ($p_d < 1$).
The numbers in the row with reference are the existing results. 
Clearly, standard DP 
and BDP with $p_d = 0.8, 0.6$ share the same critical exponents.} 
\label{tab:DP}
\end{center}
\end{table*}

The determination of the critical exponents $\delta$ and $z$ is obtained in an analogous way 
by simulating at the estimated percolation threshold, and the results are listed in Table~\ref{tab:DP}.

\begin{figure}
\includegraphics[scale=1.25]{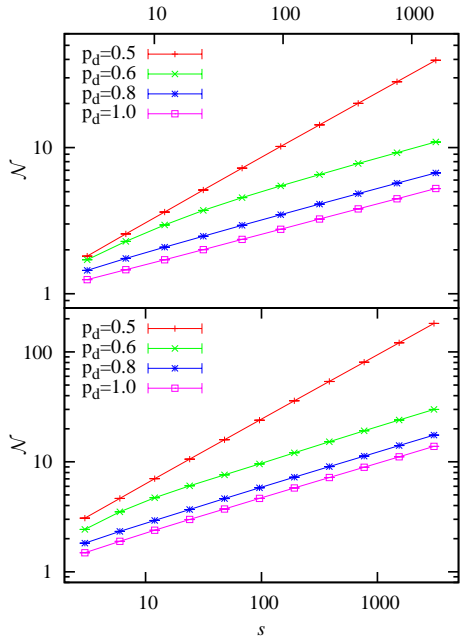}
\centering
\caption{Log-log plot of $\scrN$ versus $s$ at $p_c$.  The bottom (top) panel is for 2D (3D). 
It is clearly seen that the slope is identical for all the $p_d \neq 1/2$ cases, and is distinct 
from that of IP ($p_d=1/2$).} 
\label{fig:N}
\end{figure}

The results in Table~\ref{tab:DP} strongly suggest that, as long as $p_d$ deviates from $1/2$, 
the system falls into the standard DP universality class. 
For an illustration, we make the log-log plot of the critical quantity $\scrN$ versus the 
shell number $s$ in Fig.~\ref{fig:N}. Clearly,  the slope for $p_d=1/2$ is distinct from those for the other cases,
which are independent of $p_d$.

\section{Crossover exponent}
\begin{figure}
\includegraphics[scale=1.25]{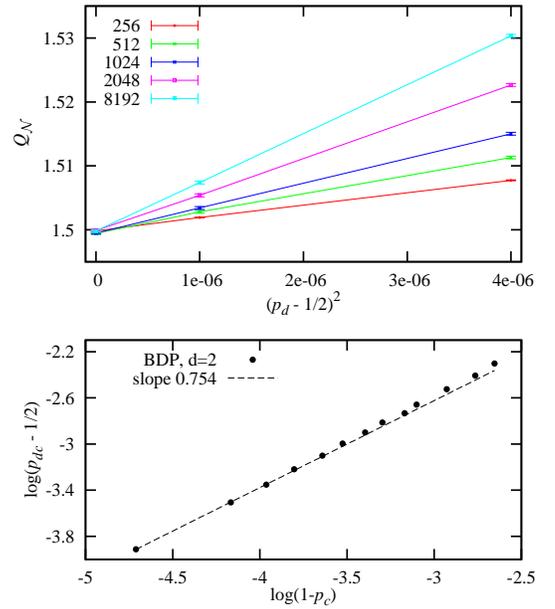}
\centering
\caption{Top: Ratio $Q_{\scrN}$ versus $(p_d - 1/2)^2$ with $p = 1$ on square lattice.
        Bottom: Log-log plot of $p_{dc}-1/2$ versus $1-p_c$ for the transition line $(p_{dc},p_c)$ near IP.
        The dashed line has slope $0.754$.}
\label{fig:fix_p}
\end{figure}

 The fact that BDP for $p_d \neq 1/2$ is in the DP universality means that in the language of renormalization group theory, 
 the operator associated with the asymmetric perturbation is relevant near the IP fixed point. 
 To confirm this, we simulate BDP near IP with $p=p_c=1$ by varying $\epsilon_d=p_d-1/2$. 
 The simulation is up to $s_{\rm max}=8192$, and $\epsilon_d$ is set at $0$, $10^{-3}$ and $2\times10^{-3}$.
 The results for $Q_{\scrN}$ in two dimensions are shown in Fig.~\ref{fig:fix_p} versus $\epsilon_d^2$;
 note that BDPs for $\pm \epsilon_d$ are identical.
 These $Q_{\scrN}$ data are also analyzed by Eq.~(\ref{eq:fitQ}) with $Y_{\epsilon}$ being replaced by 
 the exponent $Y_{\epsilon_d}$ for the symmetric scaling field 
 and the odd terms with respect to $\epsilon_d$ being set zero.
 We obtain $Y_{\epsilon_d}=0.500(5)$, which suggests that $Y_{\epsilon_d}$ may be exactly $1/2$.

According to scaling theory, the phase transition line $(p_c,p_{dc})$ approaches to the critical IP $(p_c=1,p_{dc}=1/2)$ as \cite{Riedel-1969}
 \begin{equation}
   1 - p_c \propto |(p_{dc}-1/2)|^{1/\phi} \; , 
 \label{eq:cross_exponent}
 \end{equation}
 where $\phi = Y_{\epsilon_d}/Y_{\epsilon}$ is the so-called crossover exponent.
 We carried out some Monte Carlo simulations and determined a set of critical points near IP; 
 they are 13 critical points with $p_{dc}\in [0.52,0.6]$. 
 In Fig.~\ref{fig:fix_p}, we plot $p_{dc}-1/2$ versus $1-p_c$ in log-log scale, 
 which indeed has slope approximately equal to $\phi = Y_{\epsilon_d}/Y_{\epsilon}=0.754$.

 We also perform a similar study near the critical IP in 3D, and obtain $Y_{\epsilon_d}=0.56(1)$ and $\phi=0.67(1)$.

\section{Discussion}
\begin{figure}
 \includegraphics[scale=0.9]{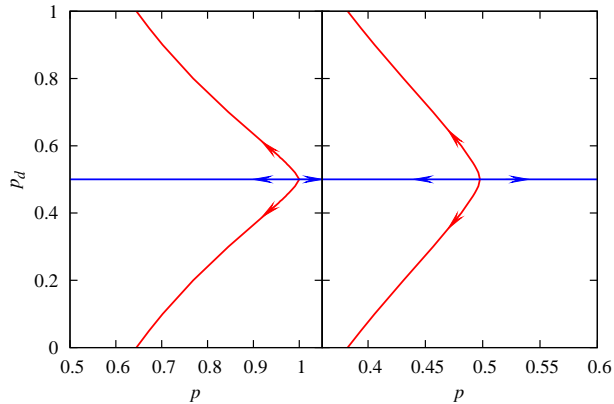}
 \centering
 \caption{Phase diagram of the BDP model in 2D (left) and 3D (right).
 The $p_d=1/2$ line corresponds to isotropic percolation. 
 The diagram for $p_d<1/2$
 is drawn by symmetry. The arrows represent the direction of the renormalization flows.}
 \label{fig:phase_diagram}
 \end{figure}

 We introduce a biased directed percolation model, which includes standard isotropic and directed 
 percolation as two special cases. Large-scale Monte Carlo simulations are carried out
 in two and three dimensions. We find that the operator associated with the anisotropy is relevant
 near the IP fixed point, which implies that BDP in the region $p_d\neq1/2$ is in the DP university class.
 On this basis, 
 the phase diagram and the associated renormalization flows are shown in Fig.~\ref{fig:phase_diagram}.
 Since the upper critical dimensionality is different for standard IP and DP,
 it is not clear whether the similar renormalization flows would hold in higher dimensions.
 We mention that such crossover phenomena have attract much attention both in the fields of 
 equilibrium and non-equilibrium statistical mechanics \cite{Pfeuty,Aharony,Lubeck-JSM,Mendes,Frojdh,Janssen,Schonmann-JSP-1986}.
 In retrospect, it is not surprising that the asymmetric perturbation is relevant near IP. 
 At IP, all the directions are equivalent and ``spatial'' and ``temporal'' directions 
 cannot be defined.  However, as soon as $p_d -1/2 \neq 0$, such a symmetry is broken and the center of the activated sites 
 moves along the ``temporal'' direction as the growing process continues.
 It is also plausible that as long as the ``spatial'' and ``temporal'' symmetry is not restored, 
 such an asymmetric perturbation is irrelevant near DP. 
 This is similar to the fact that asymmetric diffusion on the basic contact process is irrelevant~\cite{Schonmann-JSP-1986}.
 In terms of the chemical distance $s$, the effect from the anisotropy can be asymptotically 
 described as $\propto (p_d-1/2) s^{Y_{\epsilon_d}}$ with $Y_{\epsilon_d}(2D) = 0.500(5)$ and $Y_{\epsilon_d}(3D) = 0.589(10)$.
 One can also use the Euclidean distance $r$ to describe such an anisotropic effect as  $\propto (p_d-1/2) r^{1/\nu_d}$
 with $Y_{\epsilon_d}=1/(\nu_d d_{\rm min})$.
 Substituting the $d_{\rm min}$ value into $Y_{\epsilon_d} $, one obtains $\nu_d (2D) = 1.77(1)$ and  $\nu_d (3D) = 1.30(2)$.

 When viewing standard isotropic percolation in the framework of BDP,
 one observes that two independent critical exponents, e.g., $\nu$ and $\beta$, are no longer sufficient 
 to describe the critical scaling behavior. In this case, the shortest-path exponent $d_{\rm min}$ 
 appears naturally and becomes indispensable, and thus isotropic percolation also 
 has three independent critical exponents.
 Our estimate of $d_{\rm min}$ significantly improves over the existing results both in two and three dimensions.
 Our result $d_{\rm min} = 1.130\,76(10)$ does not agree with the recently
 conjectured value $217/192$ \cite{Deng-PRE-81} in two dimensions.
 This result appears to refute the conjectured value. 
 On the other hand, we note that, in terms of the chemical distance $s$, a new source of finite-size 
 corrections occurs in the scaling behavior, and these corrections are not well understood yet.
 Further, we observe that the restored symmetry for IP can be regarded as
 $\nu_\parallel = \nu_\perp$ in the BDP model. In some cases, the coincidence of two critical exponents
 may suggest the existence of logarithmic corrections of the $\log$ or $\log \log$ form, and they 
 can be either additive or multiplicative. 
 In practice, logarithmic finite-size corrections have indeed been observed 
 for standard isotropic percolation in two dimensions~\cite{Feng08}, 
 which is in terms of Euclidean distance.
 In this sense, we cannot entirely exclude 
 the possibility that the tiny difference between the present numerical result for $d_\mathrm{min}$ and the conjectured value
 arises from some unknown corrections that have not been taken into account in the numerical analysis.
 Numerical investigation of this problem seems very difficult if not impossible.
 Nevertheless, since the exact value of $d_{\rm min}$ is conjectured as a function of $q$ 
 for the $q$-state Potts model ~\cite{Deng-PRE-81}, one can accumulate more numerical evidence by studying the $q \neq 1$ case.

 Finally, the numerical estimate of the critical exponent $Y_{\epsilon_d}$ or $\nu_d$ due to 
 the asymmetric perturbation near IP raises a question: is it a ``new'' 
 independent critical exponent or simply related in some way to the known ones like $\beta$, $\nu$, and $d_{\rm min}$?
 In particular, in two dimensions, one would ask whether $\nu_d$ or $Y_{\epsilon_d}$ 
 can be exactly obtained in the framework of Stochastic Loewner Evolution (SLE), conformal field theory
 or Coulomb gas theory. 

\section{Acknowledgments}
  This work was supported in part by NSFC under Grant No. 10975127 and  91024026, and
  the Chinese Academy of Science. 
  RMZ acknowledges support from National Science Foundation Grant No.\ DMS-0553487.
  We also would like to thank Dr. Timothy M. Garoni in Monash University for valuable comments.

\end{document}